%% file: main.tex
\documentclass[%
 reprint,
nofootinbib,
 amsmath,amssymb,
 aps,
superscriptaddress,
pre,
showkeys
]{revtex4-2}

\usepackage{gensymb}
\usepackage{amsfonts}
\usepackage{tikz}
\usetikzlibrary{patterns, patterns.meta}
\usepackage{appendix}
\usepackage{standalone}
\usepackage{graphicx}
\usepackage{dcolumn}
\usepackage[caption=false]{subfig}
\usepackage{bm}
\usepackage[colorlinks=true, allcolors=blue]{hyperref}
\usepackage{cleveref}
\usepackage{comment}
\usepackage[normalem]{ulem}

\begin{document}

\preprint{APS/123-QED}

\title{Information thermodynamics of cellular ion pumps}

\author{Julián D.\ Jiménez-Paz}
 \affiliation{Physics Department, Simon Fraser University}
 \affiliation{Departamento de Física, Universidad Nacional de Colombia}
 \affiliation{Physics Department, Cornell University}
\author{Matthew P.\ Leighton}
\thanks{These authors contributed equally.}
\email{matthew.leighton@yale.edu}
\affiliation{Physics Department, Simon Fraser University}
\affiliation{Physics Department and Quantitative Biology Institute, Yale University}
\author{David A.\ Sivak}
\thanks{These authors contributed equally.}
\email{dsivak@sfu.ca}
\affiliation{Physics Department, Simon Fraser University}

\date{\today}

\begin{abstract}
The framework of bipartite stochastic thermodynamics is a powerful tool to analyze a composite system's internal thermodynamics. It has been used to study the components of different molecular machines such as ATP synthase. However, this approach has not yet been used to describe ion-transporting proteins despite their high-level functional similarity. Here we study the bipartite thermodynamics of the sodium-potassium pump in the nonequilibrium steady state. Using a physically intuitive partition between the ATP-consuming subsystem and the ion-transporting subsystem, we find considerable information flow comparable to other molecular machines, and Maxwell-demon behavior in the ATP-consuming subsystem. We vary ion concentrations and transmembrane voltage in a range including the neuronal action potential, and find that the information flow inverts during depolarization.
\end{abstract}


\maketitle

\section{\label{sec:intro}Introduction}

Molecular ion pumps are proteins whose main function is to transport ions across the cell membrane against their concentration gradients. They play an essential role in many fundamental processes for the cell such as regulation of cell volume~\cite{armstrong_nak_2003}, regulation of transmembrane voltage~\cite{xu_concept_2013,brodie_role_1987}, and thermogenesis~\cite{de_meis_role_2005}. Much research has been devoted to measuring and understanding the thermodynamic properties of their operation~\cite{slayman_electrogenic_1982,lauger_thermodynamic_1984,suzuki_microscopic_2007,lervik_thermodynamic_2012,clarke_quantitative_2013}. Important performance metrics include the work the pump does per cycle, the speed it can achieve, and its overall efficiency. Relating these metrics to underlying properties opens novel possibilities for the development of artificial molecular machines~\cite{balzani_artificial_2000,korosec_motility_2024}.

Recent progress in the theory of nonequilibrium and stochastic thermodynamics~\cite{parrondo_thermodynamics_2015,seifert_stochastic_2008,peliti_stochastic_2021,jarzynski_equalities_2011} allows us to quantify thermodynamic details in existing models for such molecular machines. In particular, the theory of bipartite thermodynamics has made it possible to resolve internal flows of energy and information between a machine's coupled components~\cite{leightonFlowEnergyInformation2025}. Recent work has sought to understand the role of information in the operation of molecular machines, with emerging evidence suggesting that paradigmatic machines such as ATP synthase~\cite{lathouwers2022internal}, photosystem II~\cite{leighton2024information}, kinesin~\cite{Takaki2022_Information,buissonHuntingMaxwellsDemon2025}, and many synthetic molecular machines~\cite{amano2022insights,corra2022kinetic} rely on information flows to transduce free energy. However, it is currently unknown whether information as a thermodynamic resource is necessary for, or even present in, the operation of cellular ion pumps.

Cells use a great variety of ion pumps for various essential purposes~\cite{dyla2020structure,kandori2015ion,sze2018plant}. Their dynamics are typically modeled by a small set of metastable states corresponding to their conformations (often just two for the P-ATPases) and the bound ions and small molecules. Here we focus on sodium-potassium pumps (also known as Na$^+$,K$^+$-ATPases) due to their extensive experimental characterization. The sodium-potassium pump is a transmembrane ion pump whose purpose is to take sodium ions out of the cell and bring potassium ions into the cell~\cite{skou_influence_1957}. It helps control the potassium concentration (important in mammals for kidney function~\cite{palmer_regulation_2015}) and helps maintain the electrochemical potential difference across the cell membrane (important in excitable cells like neurons~\cite{brodie_role_1987}).

In this paper, we use bipartite thermodynamics~\cite{ehrich_energy_2023} to quantify the efficiency, internal energy, and information flows in an experimentally validated model of the sodium-potassium pump, when varying voltage across biologically relevant ranges during the neuronal action potential. In Sec.~\ref{sec:methods} we present the relevant nonequilibrium dynamics and stochastic thermodynamics and the model for the operation of sodium-potassium pumps. In Sec.~\ref{sec:results}, we calculate the variation of probability current and internal information and energy flows across varying transition rates, ion concentrations, and transmembrane voltage. Information and heat flows between subsystems show Maxwell-demon behavior for a resting cell but not when the voltage increases. Finally, Sec.~\ref{sec:discussion} provides additional context for the obtained results, the relevance and limitations of the model, and possible future research directions.

\section{\label{sec:methods}Methods}

\subsection{Dynamics}

\begin{figure*}[ht]
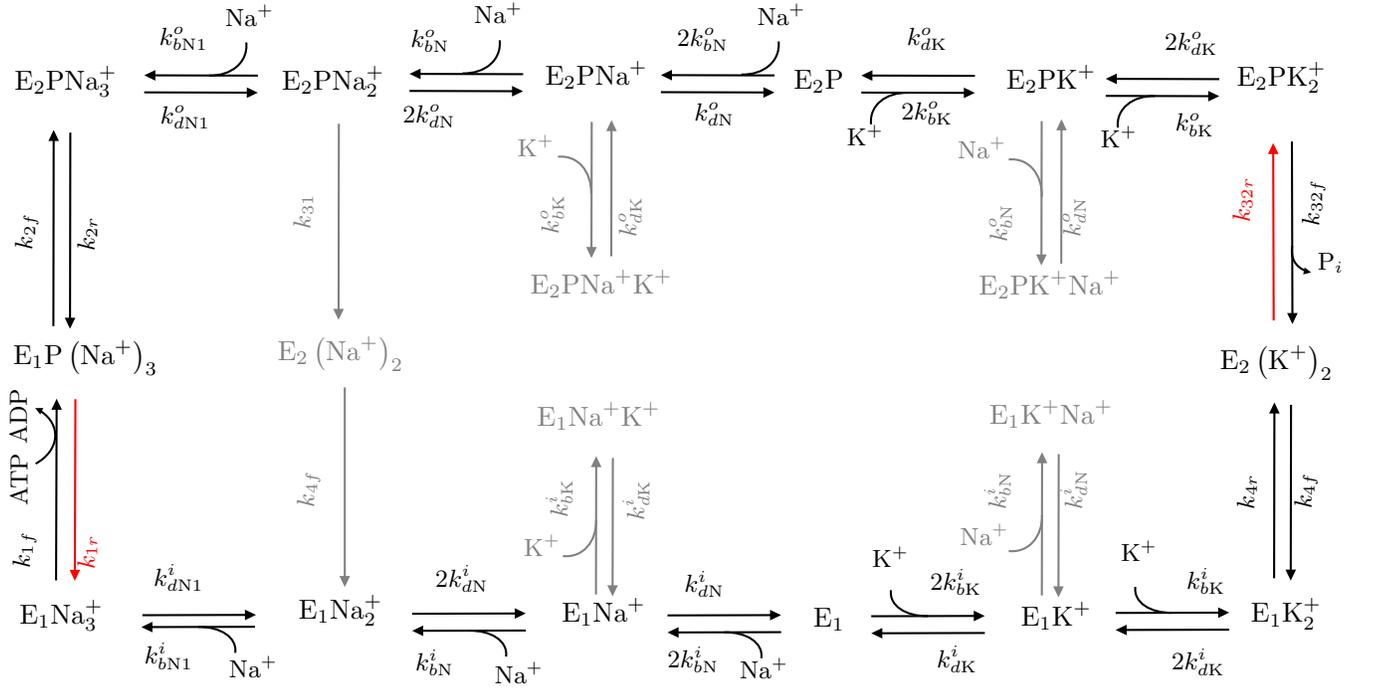

    \centering
    \includestandalone[width=\textwidth]{Albers-Post-Horizontal}
    \caption{Albers-Post cycle of the sodium-potassium pump, encompassing a main path exchanging sodium and potassium, an alternative path releasing one sodium ion to the extracellular medium, and four dead-end states corresponding to mixed ion binding. Gray: states and transitions removed in this paper; red: added reverse transitions. Adapted from~\cite{clarke_quantitative_2013}.}
    \label{fig:Albers_Post}
\end{figure*}

In a given cycle, the sodium-potassium pump moves three sodium ions out and two potassium ions into the cell, both against their concentration gradients. In doing so, it hydrolyzes one ATP into ADP and inorganic phosphate. The sodium-potassium pump has two conformational states E1 and E2 that define the binding-site exposure of the protein. In the E1 state, the protein exposes its ion-binding sites to the intracellular medium and has high affinity for sodium ions. Once it has bound three sodium ions, the protein is phosphorylated, conformationally switching to E2P where the ion-binding sites are exposed to the extracellular medium. (P refers to the attached phosphate). In its phosphorylated state, the sodium-potassium pump has higher affinity for potassium ions and lower affinity for sodium ions; therefore, it binds two potassium ions after releasing the sodium ions to the extracellular medium. In return, this triggers dephosphorylation and conformational switching back to the E1 state. The Albers-Post cycle~\cite{albers_biochemical_1967, post_activation_1972} (Fig.~\ref{fig:Albers_Post}) is a detailed description of the sodium-potassium pump operation. It accounts for the transition rates between states along both the main pathway just described and an alternative path where the E2P state, with two sodium ions bound, dephosphorylates and switches to E1, thus on net only transporting one sodium ion out of the cell. The Albers-Post cycle also accounts for four `dead-end' states where the protein simultaneously binds a sodium and a potassium ion. 

The evolution of the probability distribution over the states in the Albers-Post model can be described through a master equation
\begin{equation}
    \dot{\mathbf{P}} = \mathbf{RP} \ ,
\end{equation}
where $\mathbf{P}$ is the probability vector for the system's states and $\mathbf{R}$ is the transition matrix whose entries $R_{zz'}$ quantify the respective transition rates from state $z'$ to state $z$. The steady-state distribution $\mathbf{P}_\text{NESS}$ solves the equation
\begin{equation}
    \mathbf{RP}_\text{NESS} = 0 \ ,
\end{equation}
i.e., is the eigenvector (normalized to one) with eigenvalue zero. We solve this equation using the C++ linear-algebra library Eigen~\cite{guennebaud_eigen_2010}; our code is publicly available on Github~\cite{github}.

The steady-state probability current flowing through the main (Na$^+$/K$^+$) path is over 300$\times$ that through the alternative (Na$^+$/Na$^+$) path~\cite{clarke_quantitative_2013}, so we remove the alternative path (through state E$_2$(Na$^+$)$_2$) to focus on the main path. We also remove the dead-end states (E$_1$Na$^+$K$^+$, E$_1$K$^+$Na$^+$, E$_2$PNa$^+$K$^+$, and E$_2$PK$^+$Na$^+$) that do not contribute to the steady-state flux. 

For thermodynamic consistency, each transition with positive rate should have a corresponding reverse transition with positive rate. Therefore, we add to the (pruned) Albers-Post model the transition rate $R_{01}=k_{1r}$ between states E$_1$Na$^+_3$ and E$_1$P(Na$^+$)$_3$, and the transition rate $R_{78}=k_{32r}$ between states E$_2$PK$^+_2$ and E$_2$(K$^+$)$_2$. Figure~\ref{fig:Albers_Post} shows the resulting scheme, with 0 the bottom left state and state numbering increasing clockwise around the cycle.

\subsection{Bipartite thermodynamics}

Molecular pumps operate in a nonequilibrium steady state (NESS) where the probability $P(z)$ of being in a given state $z$ is constant in time, and the current
\begin{equation}
    J_{zz'} \equiv R_{zz'}P(z') - R_{z'z}P(z)
\end{equation}
from state $z'$ to state $z$ is generally nonzero.

Such a nonequilibrium steady state supports constant rates of work $\dot{W}$ done on the system, and of heat $\dot{Q}$ flowing from the environment into the system. These quantities are related by the first law of thermodynamics, with constant internal energy ($\dot{E}=0$) at steady state:
\begin{equation}\label{eq:first_law}
    \dot{W}+\dot{Q} = 0 \ .
\end{equation}
Work can be divided into chemical work $\dot{W}_\text{chem}$ related to energy exchange with chemical reservoirs and electrical work $\dot{W}_\text{el}$ due to the transmembrane electric voltage. The chemical work rate is~\cite{ehrich_energy_2023}
\begin{equation}\label{eq:W_chem_definition}
    \dot{W}^\text{chem} \equiv - \sum_{z>z'}[R_{zz'}P(z')-R_{z'z}P(z)]\Delta \mu_{zz'} \ ,
\end{equation}
where $\Delta \mu_{zz'}$ is the free energy difference in the environment when transitioning from state $z'$ to $z$. The transmembrane electric voltage is $V=-79$mV (potential inside the cell minus that outside the cell)~\cite{clarke_quantitative_2013}. The electrical work rate is thus
\begin{equation}\label{eq:W_mech_definition}
\dot{W}^\text{el} \equiv - \sum_{z>z'}[R_{zz'}P(z')-R_{z'z}P(z)]\frac{V}{2} q_{zz'}\ ,
\end{equation}
where $q_{zz'}$ is the transported charge from outside to inside, equal to either $0$, $e$, or $-e$, where $e$ is the fundamental electric charge. Lacking detailed information about states involving inaccessible ions (middle row in Fig.~\ref{fig:Albers_Post}), we evenly split the voltage difference between extracellular and intracellular-facing pump states.

The unicyclic topology of the model imposes a single steady-state current $J=J_{zz'}$ for all $z>z'$, so we can split the chemical work $\dot{W}^\mathrm{chem}$ into contributions from the different chemical species and simplify as
\begin{subequations}\label{chem_work_decomp}
    \begin{align}
        \dot{W}^{\text{chem, K}^+} &\equiv -2\Delta \mu_{\mathrm{K}^+} J \ , \\
        \dot{W}^{\text{chem, Na}^+} &\equiv -3 \Delta \mu_{\mathrm{Na}^+} J\ ,\\
        \dot{W}^{\text{chem, ATP}} &\equiv -\Delta\mu_\mathrm{ATP} J\ .
    \end{align}
\end{subequations}
Here $\Delta \mu_{\mathrm{K}^+} \equiv k_\text{B}T \ln\,[\text{K}^+]^{\rm i}/[\text{K}^+]^{\rm o} $, $\Delta \mu_{\mathrm{Na}^+} \equiv k_\text{B}T \ln\,[\text{Na}^+]^{\rm o}/[\text{Na}^+]^{\rm i}$, and $\Delta \mu_{\mathrm{ATP}} \equiv k_\text{B}T \ln\,[\mathrm{ADP}][\mathrm{P_i}]/[\mathrm{ATP}]K_\mathrm{h}$, with superscripts i and o respectively referring to concentrations \textit{inside} and \textit{outside} of the cell. The integer prefactors in Eqs.~\eqref{chem_work_decomp} quantify the number of ions (Na$^+$ and K$^+$ respectively) transported per cycle.

We determine the efficiency as
\begin{equation}
    \eta 
    \equiv
    \frac{-\dot{W}^{\text{chem, K}^+} - \dot{W}^{\text{chem, Na}^+} - \dot{W}^\text{el}}{\dot{W}^{\text{chem, ATP}}} \ ,
\end{equation}
the ratio of the output work from the system to the environment (in this case from ion pumping) and the input work from the environment to the system (here from ATP hydrolysis).

In the previous subsection we introduced additional reverse transition rates into the Albers-Post model to ensure thermodynamic consistency. Unlike the other rates, these have not been determined experimentally in the literature. Thus, we constrain the product of the introduced transition rates by considering the affinity (total scaled free energy change)~\cite{brown_theory_2020} over the cycle,
\begin{equation}
    \beta W = \ln \frac{\prod_{i=1}^N R_{ij}}{\prod_{i=1}^NR_{ji}} \ ,
\end{equation}
where $\beta \equiv (k_\text{B}T)^{-1}$, $k_\text{B}$ is the Boltzmann constant, $T$ is the temperature, and $W$ is the total work done per cycle. We use reported values~\cite{clarke_quantitative_2013} for the original transition rates, and for the introduced transition rates we obtain the relation
\begin{equation}
    R_{0,1}R_{7,8} = \left(\prod_{i>j} R_{ij}\right)\left(\prod_{\substack{i>j\\ i\neq0,7}}R_{ji}\right)^{-1} \exp(-\beta W) \ ,
\end{equation}
leaving one free parameter $\gamma = \frac{R_{0,1}}{R_{7,8}}$. This parameter quantifies the proportion of the reverse transition rate for phosphorylation (from state E$_1$P(Na$^+$)$_3$ to state E$_1$Na$^+_3$) to the reverse transition rate for dephosphorylation (from state E$_2$(K$^+$)$_2$ to state E$_2$PK$^+_2$).

The heat flow is 
\begin{equation}\label{eq:heat_definition}
    \dot{Q} \equiv -k_\text{B}T \sum_{z>z'}[R_{zz'}P(z')-R_{z'z}P(z)]\ln\frac{R_{zz'}}{R_{z'z}} \ ,
\end{equation}
Finally, the rate of change of system internal energy is
\begin{equation}\label{eq:energy_definition}
    \dot{E} \equiv \sum_{z>z'}[R_{zz'}P(z')-R_{z'z}P(z)](\epsilon_z - \epsilon_{z'}) \ ,
\end{equation}
where $\epsilon_z - \epsilon_{z'}$ is the difference in internal energy between states $z'$ and $z$. Note that the state-dependent internal energy $\epsilon_z$ can also be interpreted as an internal free energy since the state $z$ is in reality a mesostate coarse-graining many microstates~\cite{espositoStochasticThermodynamicsCoarse2012}. Similarly $\dot{E}$ could be interpreted as the rate of change of the internal free energy. Nevertheless, we will refer to $\epsilon_z$ and $E$ as internal energies to avoid confusion with the transduced free energy (defined at the end of this section).

The rates $R_{zz'}$ and $R_{z'z}$ satisfy a local detailed-balance relation
\begin{equation}\label{eq:local_detailed_balance}
\ln \frac{R_{zz'}}{R_{z'z}} = \frac{\epsilon_{z'}-\epsilon_{z}-\Delta \mu_{zz'}-q_{zz'}V/2}{k_\text{B}T} \ .
\end{equation}

Some machines are composed of coupled components that can be treated as distinct thermodynamic systems. If we can divide a system $Z$ into two subsystems $X$ and $Y$ with degrees of freedom subject to independent fluctuations, we call these \textit{bipartite} subsystems~\cite{leightonFlowEnergyInformation2025,ehrich_energy_2023}. For discrete degrees of freedom, as in our case, this means the subsystems do not change states simultaneously. Thus, we can decompose the transition rates as
\begin{equation}
    R_{zz'} = R^{xx'}_y\,\delta_{yy'} + R_{yy'}^x\,\delta_{xx'} \ ,
\end{equation}
for Kronecker delta $\delta$.

The bipartite assumption permits subsystem-specific energy balances in terms of subsystem-specific currents 
\begin{subequations}
    \begin{align}
        J^{xx'}_y &\equiv R^{xx'}_yP(x',y)-R^{x'x}_yP(x,y) \\
        J^x_{yy'} &\equiv R^x_{yy'}P(x,y')-R^x_{y'y}P(x,y) \ ,
    \end{align}
\end{subequations}
and subsystem decompositions of heat flows
\begin{subequations}\label{eq:subsystem_Q}
    \begin{align}
        \dot{Q} &\equiv \dot{Q}^X + \dot{Q}^Y \\
         \dot{Q}_X &\equiv -k_\text{B}T \sum_{x>x',y}J^{xx'}_y \ln \frac{R^{xx'}_y}{R^{x'x}_y} \\
         \dot{Q}_Y &\equiv -k_\text{B}T \sum_{x,y>y'} J^x_{yy'} \ln \frac{R^x_{yy'}}{R^x_{y'y}} \ ,
    \end{align}
\end{subequations}
chemical work rates
\begin{subequations}\label{eq:subsystem_W_chem}
    \begin{align}
        \dot{W}^\text{chem} &= \dot{W}^\text{chem}_X +  \dot{W}^\text{chem}_Y \\
        \dot{W}^\text{chem}_X &\equiv -\sum_{x>x',y} J^{xx'}_y  \Delta \mu^{xx'}_y \\
        \dot{W}^\text{chem}_Y &\equiv -\sum_{x,y>y'}J^x_{yy'} \Delta \mu ^x_{yy'} \ ,
    \end{align}
\end{subequations}
electrical work rates
\begin{subequations}\label{eq:subsystem_W_mech}
    \begin{align}
        \dot{W}^\text{el} &= \dot{W}^\text{el}_X +  \dot{W}^\text{el}_Y \\
        \dot{W}^\text{el}_X &\equiv -\sum_{x>x',y} J^{xx'}_y \frac{V}{2}q^y_{xx'} \\
        \dot{W}^\text{el}_Y &\equiv -\sum_{x,y>y'}J^x_{yy'} \frac{V}{2}q^x_{yy'} \ ,
    \end{align}
\end{subequations}
and internal energy changes
\begin{subequations}\label{eq:subsystem_energy}
    \begin{align}
        \dot{E} &= \dot{E}_X+ \dot{E}_Y \\
        \dot{E}_X &\equiv \sum_{x>x',y} J^{xx'}_y (\epsilon_{xy}-\epsilon_{x'y}) \\
        \dot{E}_Y &\equiv \sum_{x,y>y'}J^x_{yy'} (\epsilon_{xy}-\epsilon_{xy'}) \ .
    \end{align}
\end{subequations}

A complete thermodynamic accounting requires also quantifying information, a thermodynamic resource that, loosely speaking, captures the correlation between different subsystems. Information flow from one subsystem permits the other subsystem to convert heat into work, in apparent violation of the second law~\cite{leightonFlowEnergyInformation2025}.

Information between random variables $X$ and $Y$ is quantified by the mutual information $I[X,Y]$~\cite{cover1999elements}. At steady state the time derivative of the mutual information 
vanishes: 
$\mathrm{d}_tI=0$. It can however be split into contributions from each subsystem, which we call information flows $\dot{I}_X$ and $\dot{I}_Y$. The information flow $\dot{I}_X$ is the rate at which the dynamics of the $X$ subsystem increase the mutual information, defined formally as~\cite{leightonFlowEnergyInformation2025}
\begin{equation}\label{eq:info_flow_definition}
\dot{I}_X \equiv \lim_{\tau\to0}\frac{I\left[X(t+\tau);Y(t)\right] - I\left[X(t);Y(t)\right]}{\tau}.
\end{equation}
The information flow $\dot{I}_Y$ is defined analogously. Since at steady state the mutual information is constant, the two flows satisfy
\begin{equation}\label{eq:info_constraint}
    \dot{I}^X = -\dot{I}^Y.
\end{equation}
In terms of the master-equation dynamics considered in this paper, the information flows can be computed as~\cite{ehrich_energy_2023}
\begin{subequations}\label{eq:subsystem_information}
    \begin{align}  
        \dot{I}_X &\equiv \sum_{x>x',y} J^{xx'}_y\ln\frac{P(y|x)}{P(y|x')} \\
        \dot{I}_Y &\equiv \sum_{x,y>y'}J^x_{yy'} \ln\frac{P(x|y)}{P(x'|y)}
    \end{align}
\end{subequations}
due to the respective dynamics of subsystems $X$ and $Y$. Here $P(y|x)$ is the conditional probability of subsystem $Y$ being in state $y$ given that subsystem $X$ is in state $x$~\cite{ehrich_energy_2023}, and information flow is measured in nats/s. Each subsystem information flow quantifies how the dynamics of that subsystem increase the mutual information, such as in Eq.~\eqref{eq:info_flow_definition}. Information-theoretically, mutual information quantifies how knowledge about one subsystem would decrease uncertainty in the other; thermodynamically, mutual information (as a measure of reduced joint entropy) is a free-energetic resource~\cite{parrondo_thermodynamics_2015}.

{We analyze the change in each thermodynamic quantity per cycle ($\Delta E$, $\Delta I$, $W$, and $Q$) and the behavior of the probability current.

The sum $\dot{E}_X +  k_\mathrm{B}T\dot{I}_X$ is the transduced free energy from $X$ to $Y$, with $\dot{E}_X$ and $\dot{I}_X$ respectively constituting the energetic and entropic contributions~\cite{barato2017thermodynamic,lathouwers2022internal}. Therefore, a natural definition of each subsystem's efficiency is~\cite{leightonSubsystemEfficiencies}
\begin{subequations}
    \begin{align}
        \eta_X = \frac{\dot{E}_X + k_\text{B}T\dot{I}_X}{\dot{W}_X} \label{eq:eff_X} \\
        \eta_Y = \frac{-\dot{W}_Y}{\dot{E}_X + k_\text{B}T\dot{I}_X} \ . \label{eq:eff_Y}
    \end{align}
\end{subequations}
The product of subsystem efficiencies is the overall efficiency of the whole system,
\begin{equation}
    \eta = \eta_X \eta_Y \ .
\end{equation}

\subsection{Bipartite model for the sodium-potassium pump}

We propose a natural bipartite partition of the system transitions into separate transitions of two subsystems $X$ and $Y$, where subsystem $X$ captures the conformational changes in the protein (and therefore its binding-site exposure) while subsystem $Y$ is defined by the ions bound to the protein (Fig.~\ref{fig:bipartite_structure1}):
\begin{subequations}
    \begin{align}
        X &\equiv \{\text{E1}, \text{E1P}, \text{E2}, \text{E2P}\} \\
        Y &\equiv \{\varnothing,\text{Na}^+, \text{Na}^+_2, \text{Na}^+_3, \text{K}^+, \text{K}^+_2\} \ ,
    \end{align}
\end{subequations}
where $\varnothing$ means that there are no ions bound (i.e., $P(Z=\text{E1})=P(X=\text{E1},Y=\varnothing)$ and $P(Z=\text{E2P}) = P(X=\text{E2P},Y=\varnothing)$).

\begin{figure}[h]
    \centering
    \begin{subfloat}[\label{fig:bipartite_structure1}]
        {\includegraphics[width=\linewidth]{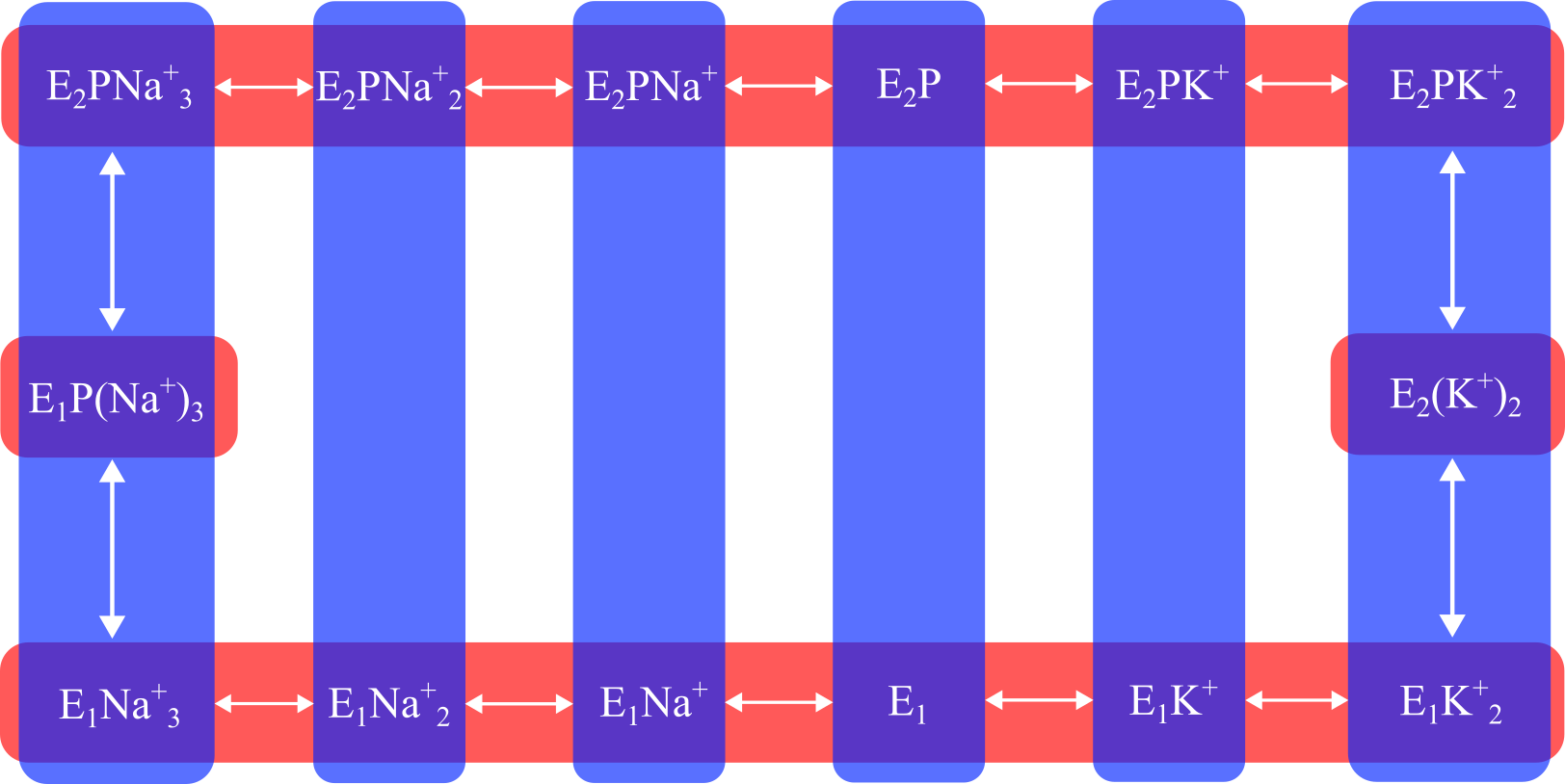}}
    \end{subfloat}
    \begin{subfloat}[\label{fig:bipartite_structure2}]
        {\includegraphics[width=\linewidth]{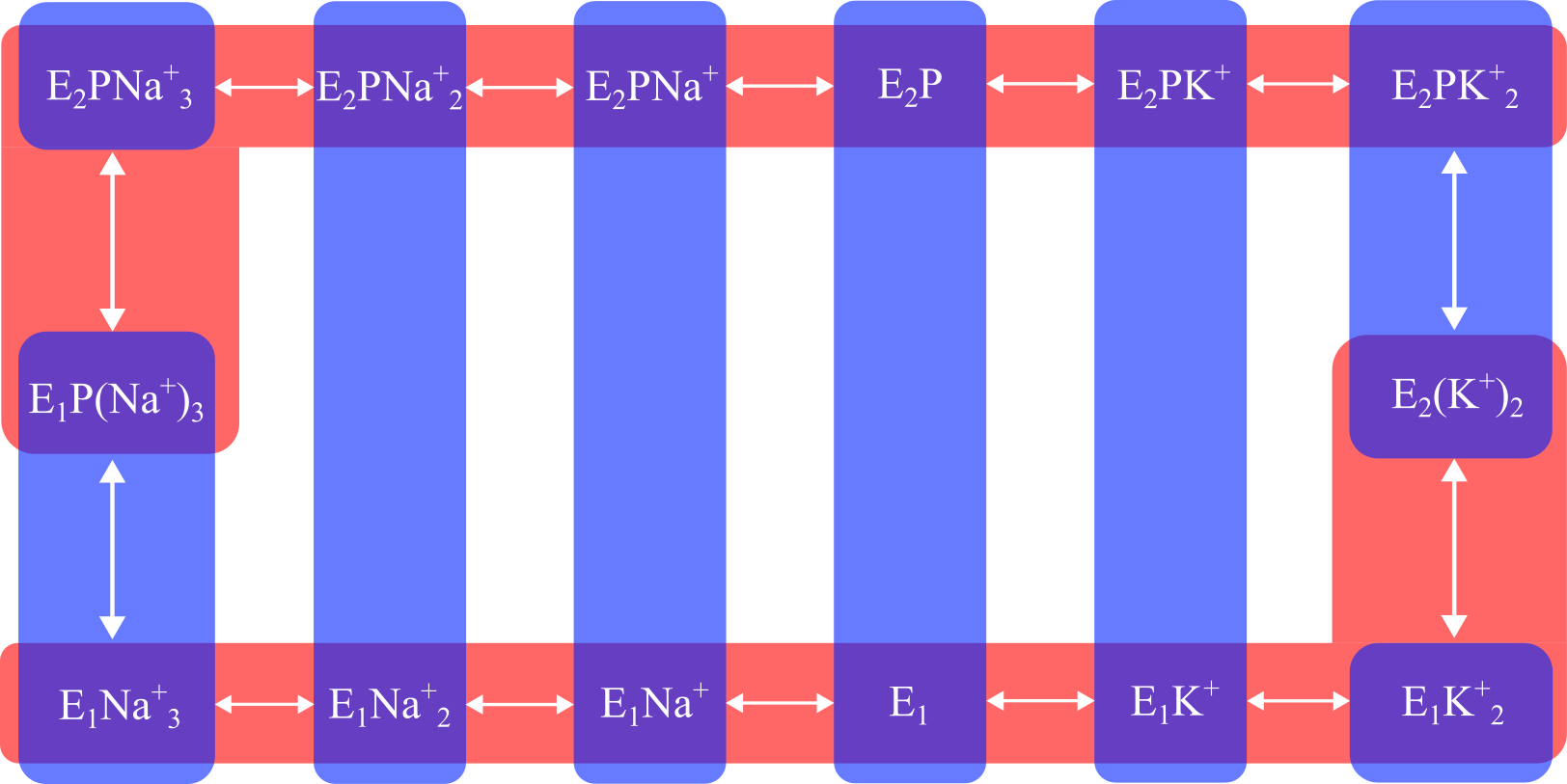}}
    \end{subfloat}
    \caption{Two different bipartite partitions. Red blocks: subsystem $X$ states; blue blocks subsystem $Y$ states. a) The principal partition in this paper. b) An alternative partition to assess the robustness of our results.}
\end{figure}

This partition is also inspired by the thermodynamics of the molecular machine, i.e., changes in the state of subsystem $X$ can be associated with ATP hydrolysis and phosphorylation while changes in the state of the $Y$ subsystem refer to work done by the protein in transporting ions. In other words, transitions in the $X$ subsystem exchange energy with the external reservoir of ATP, ADP, and P$_{\rm i}$; transitions in the $Y$ subsystem exchange energy with the external reservoir of ions. For every transition, only one subsystem ($X$ or $Y$) changes state. Thus, this partition satisfies the bipartite assumption.

Additionally, as stated before, in a steady state mutual information is constant; however, there still is information flow between subsystems constrained by Eq.~\eqref{eq:info_constraint}. Information flow between these subsystems indicates that the mutual information is continuously increased by the dynamics of one subsystem, and decreased by the dynamics of the other. This reflects transition rates of one subsystem (e.g., $X$) that strongly depend on the state of the other subsystem (e.g., $Y$), such that $X$ can be thought of as ``measuring" the state of $Y$.

\section{\label{sec:results}Results}

We first confirmed that our master-equation approach yielded the same steady-state distribution as~\cite{clarke_quantitative_2013}. 

Turning to analysis of this pruned model (Fig.~\ref{fig:Albers_Post}) for concentrations from~\cite{clarke_quantitative_2013} [App.~\ref{app:concentrations}], we varied the reverse-rate ratio $\gamma$ across several orders of magnitude. The probability current tends to zero when the two coefficients differ greatly, and it is high when they are comparable, specifically reaching a peak near $\gamma = 0.09$ (Fig.~\ref{fig:prop_turnover}).

\begin{figure}[h]
    \centering
    \begin{subfloat}[\empty\label{fig:prop_turnover}]
        {\includegraphics[width=\linewidth]{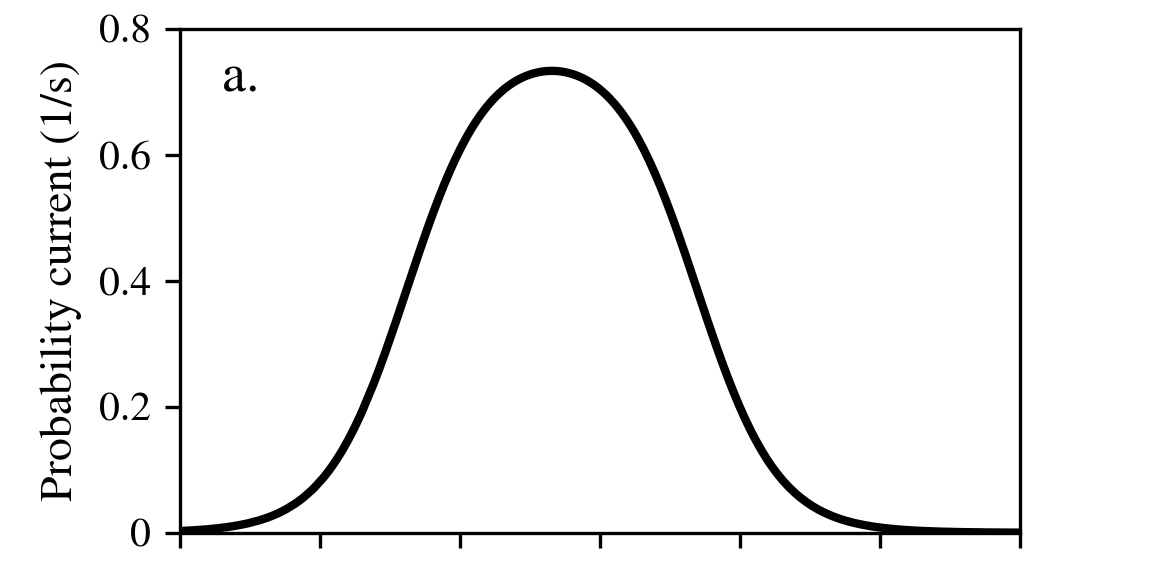}}
    \end{subfloat}
    \begin{subfloat}[\empty\label{fig:prop_energy_info}]
        {\includegraphics[width=\linewidth]{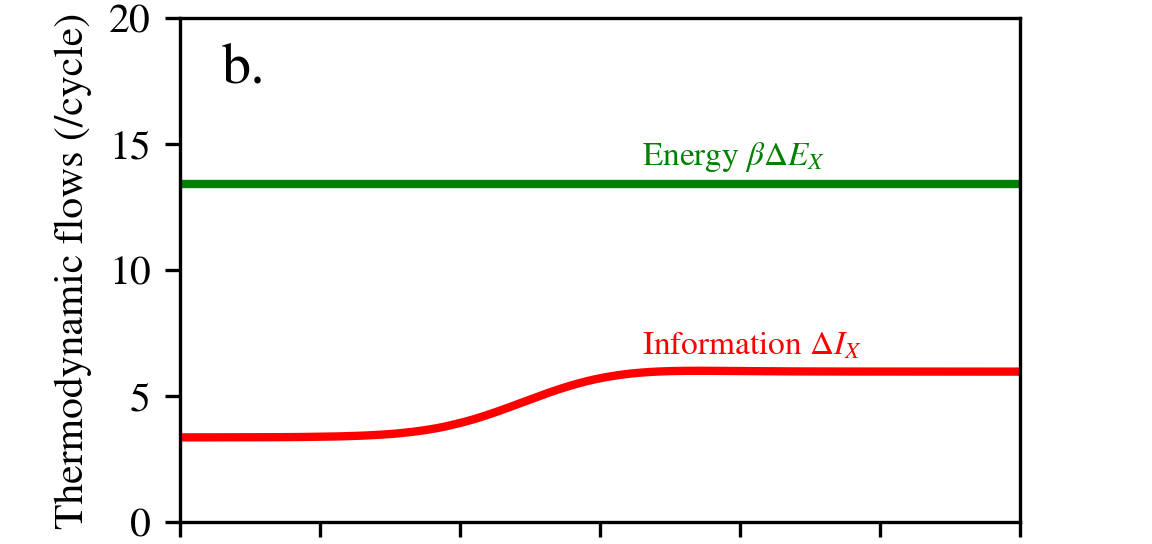}}
    \end{subfloat}
    \begin{subfloat}[\empty\label{fig:prop_efficiencies}]
        {\includegraphics[width=\linewidth]{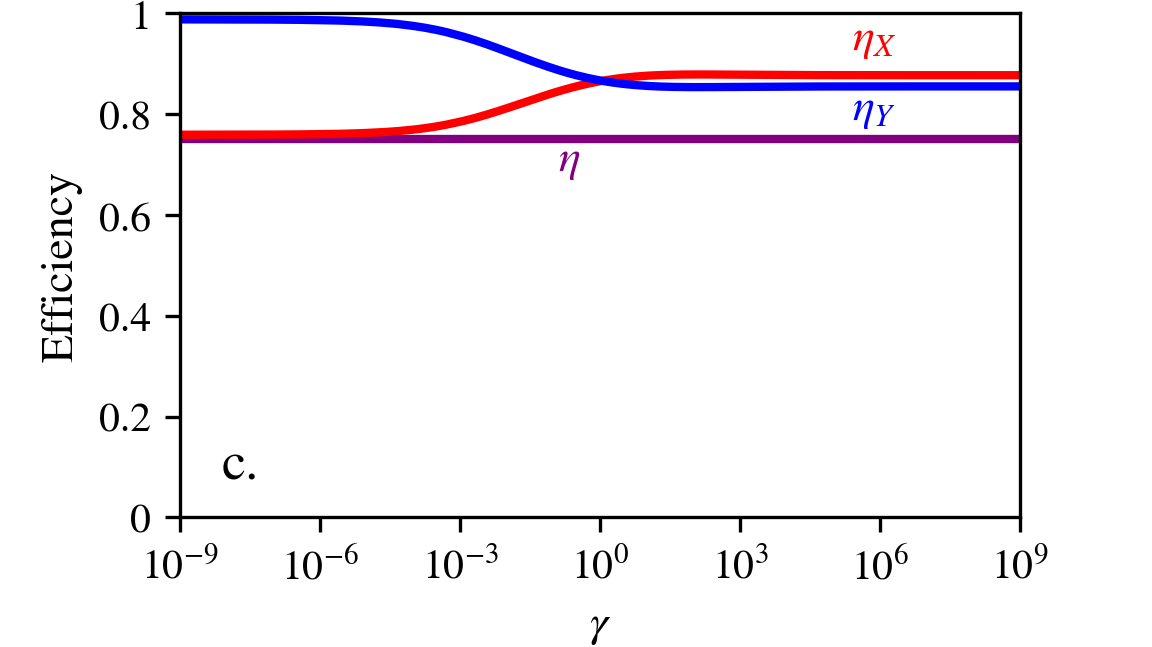}}
    \end{subfloat}
    \caption{Global probability current (a), internal energy ($k_\text{B}T/$cycle) and information (nats/cycle) flows from subsystem $X$ to subsystem $Y$ (b), and subsystem efficiencies (c) as a function of the ratio $\gamma \equiv \frac{R_{0,1}}{R_{7,8}}$ of introduced reverse rates. Environmental parameters taken from Table~\ref{tab:concentrations}.}
    \label{fig:turnover_energy_info}
\end{figure}

Figure~\ref{fig:prop_energy_info} shows the internal energy and information flows per cycle, $\Delta E_X = \dot{E}_X / J_{xx'}$ and $\Delta I_X = \dot{I}_X / J_{xx'}$. Internal energy flow $\Delta E_X \approx 13 \, k_\text{B}T / \text{cycle}$ is independent of the ratio $\gamma$ of added reverse rates. Information flow also tends to be independent of $\gamma$ for $R_{0,1} \ll R_{7,8}$ ($\gamma \ll 10^{-3}$), increases with increasing $\gamma$ in the interval $10^{-3} \lesssim \gamma \lesssim 1$, and is again independent for $R_{0,1} \gg R_{7,8}$ ($\gamma \gg 1$). Additionally, both global work and heat flows per cycle are independent of $\gamma$ (not shown). The subsystem heat flows per cycle are $\beta Q_X = -8.69$ and $\beta Q_Y = 3.17$. These values indicate that input work from ATP hydrolysis (into the $X$ subsystem) drives output work transporting ions against electrochemical potential gradients (from the $Y$ subsystem). The proportion $\frac{\dot{I}_X}{\beta\dot{E}_X + \dot{I}_X}$ of information flow to transduced free energy is important, ranging from $0.20$ to $0.31$. Therefore, a significant part (20-30\%) of the free energy transduced from X to Y is in the form of information flow.

Figure~\ref{fig:prop_efficiencies} shows the efficiency of the ion pump. Since $\Delta E_X$ is constant, the efficiency of subsystem $X$ scales linearly with the information flow $\Delta I_X$ according to Eq.~\eqref{eq:eff_X}. When $R_{0,1} \ll R_{7,8}$, the efficiency of subsystem $Y$ tends to unity, and thus $\eta \approx \eta_X$. This highlights another advantage of the bipartite decomposition: it permits decomposition of the overall efficiency into subsystem efficiencies, giving a finer-grained picture of free-energy flow within the system~\cite{leightonSubsystemEfficiencies}.

Notably, the heat in the (ion-transporting) $Y$ subsystem is positive, i.e., heat flows in. This indicates that the (ATP-consuming) $X$ subsystem acts as a Maxwell demon~\cite{lutz_information_2015}, i.e., $X$ builds up information that $Y$ then uses as a resource to convert heat into work. We can interpret the ATP subsystem as ``locking in'' favorable fluctuations of ions binding to the protein by quickly changing the protein's conformational state, which comes at the cost of energy dissipation in the form of heat released by ATP hydrolysis.

We also considered the only other partition for this specific model where ATP hydrolysis and ion binding/unbinding processes belong to different subsystems (Fig.~\ref{fig:bipartite_structure2}). For $\gamma = 1$, the information flow is $\Delta I_X =8.77$ nats/cycle and the energy flow is $\beta \Delta E_X = 10.41$ per cycle; thus the information flow is approximately $45\%$ of the total transduced free energy. For this decomposition the heat flows per cycle are $\beta \Delta Q_X = -11.70$ and $\beta \Delta Q_Y = 6.18$: there is still heat flowing into subsystem $Y$. In summary, even for the other biophysically motivated partition of this model the information flow remains a considerable proportion of the transduced free energy and the system still operates as a Maxwell demon.

\subsection{Voltage variation in neuronal depolarization}

The sodium-potassium pump is fundamental for recovering the resting transmembrane voltage after the neuronal action potential~\cite{glynn2002hundred}. When a neuron is at rest, it maintains a negative electric potential across the cell membrane. If an outside impulse is sufficiently large to raise the voltage past its threshold (usually around $-55$\,mV)~\cite{beanActionPotentialMammalian2007}, the voltage-gated sodium ion channels open, letting sodium ions passively diffuse into the cell, increasing the electric potential until it is no longer negative (known as the depolarization phase) and reaching a maximum that depends on the type of neuron. After the neuron's depolarization, the sodium channels inactivate and the potassium channels open (the absolute refractory period), letting potassium ions out of the cell and starting the repolarization phase. Then, the sodium channels close as the voltage decreases (the relative refractory period). During this process, voltage may `overshoot', giving rise to a hyperpolarization phase before the resting potential is restored~\cite{hodgkinQuantitativeDescriptionMembrane1952, purvesVoltageGatedIonChannels2001}.

\begin{figure}[h]
    \centering
    \includegraphics[width=\linewidth]{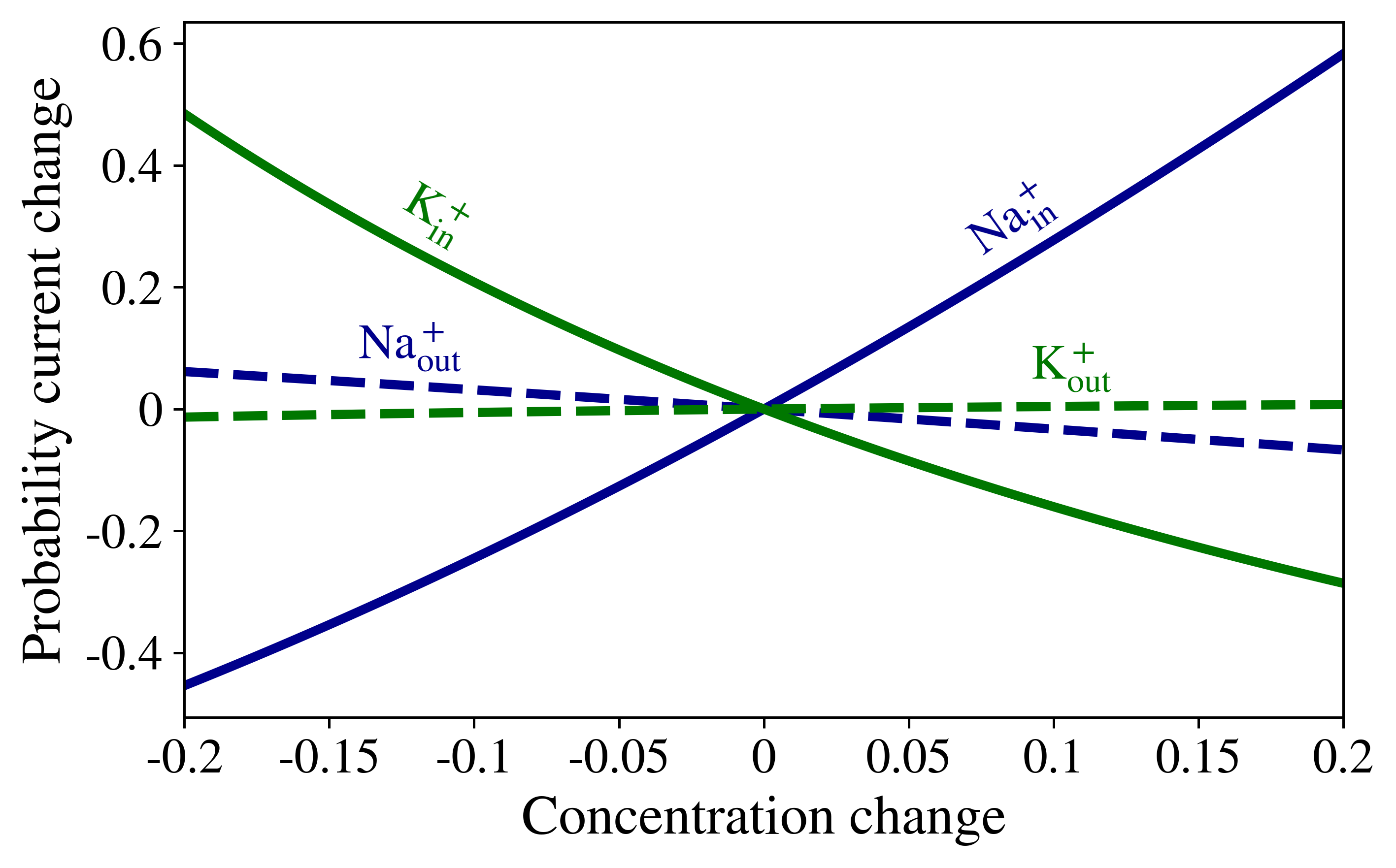}
    \caption{Relative change $(P-P_0)/P_0$ in probability current as a function of the relative change $([X^+]-[X^+]_0)/[X^+]_0$ in concentration of various ions $X^+$. $\gamma=1$ throughout. Environmental parameters taken from Table~\ref{tab:concentrations_neuron}.}
    \label{fig:concentrations_variation}
\end{figure}

The sodium-potassium pump is working throughout the entire cycle, and its operation varies due to the change in external parameters.
For a spherical cell with a membrane capacitance of 1 \textmu F/cm$^2$ \cite{golowasch_membrane_2009}, the changes in sodium and potassium ion concentrations during depolarization is $0.004\%$ and $0.06\%$ respectively. Figure~\ref{fig:concentrations_variation} shows how the probability current behaves when varying individual ion concentrations away from those of a resting neuron (App.~\ref{app:concentrations}). Changes in external ion concentrations do not significantly affect the probability current, while changes in internal ions do, with internal sodium concentration having the strongest effect. Overall, changes in ion concentrations during depolarization are too small to significantly affect the thermodynamic operation of the pump. Therefore, we focus on voltage variation.

To define a physiologically significant voltage range we estimate the equilibrium electrochemical potential for sodium and potassium using the Nernst equation
\begin{equation}
    V_\text{eq} = \frac{RT}{qF}\ln \frac{[A]_\text{out}}{[A]_\text{in}} \ ,
\end{equation}
for ideal gas constant $R$, Faraday constant $F$, and ion electric charge $q$. The resulting equilibrium potentials are $-87 \, \mathrm{mV}$ for potassium and $60\,\mathrm{mV}$ for sodium.

\begin{figure}[h]
    \centering
    \begin{subfloat}[\empty\label{fig:voltage_turnover}]
        {\includegraphics[width=\linewidth]{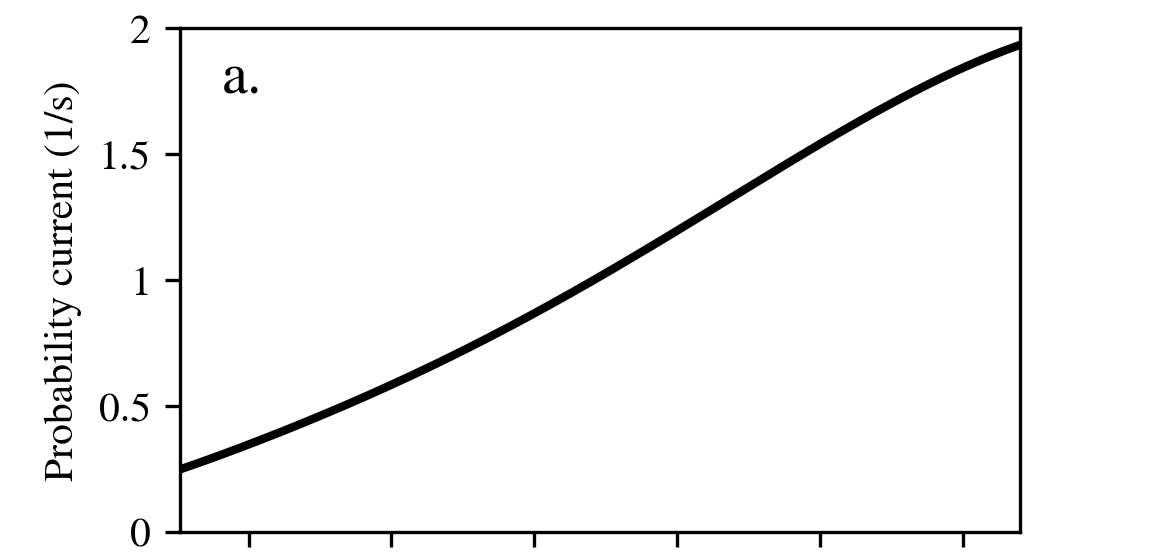}}
    \end{subfloat}
    \begin{subfloat}[\empty\label{fig:voltage_energy_info}]
        {\includegraphics[width=\linewidth]{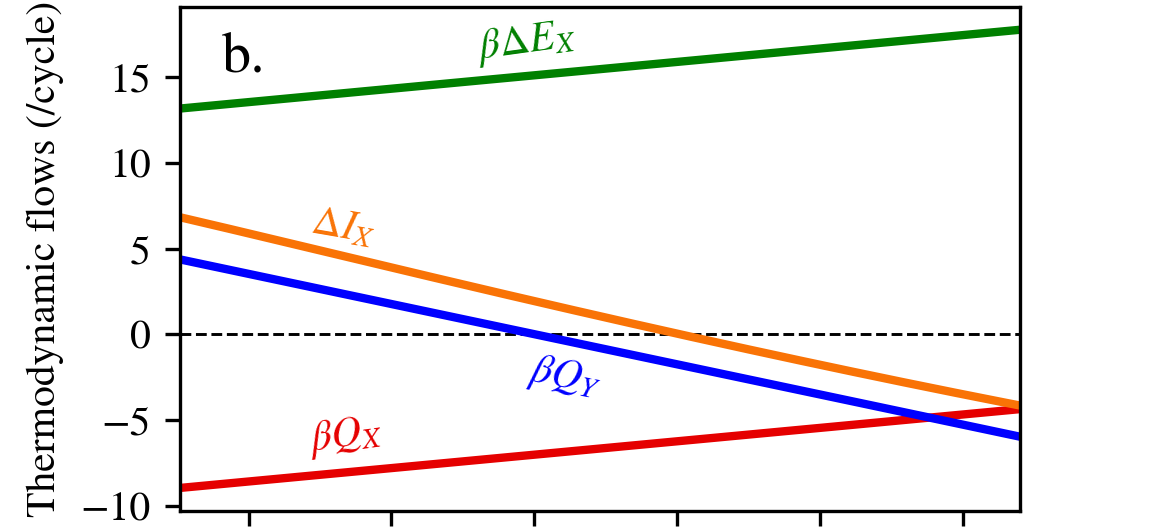}}
    \end{subfloat}
    \begin{subfloat}[\empty\label{fig:voltage_efficiencies}]
        {\includegraphics[width=\linewidth]{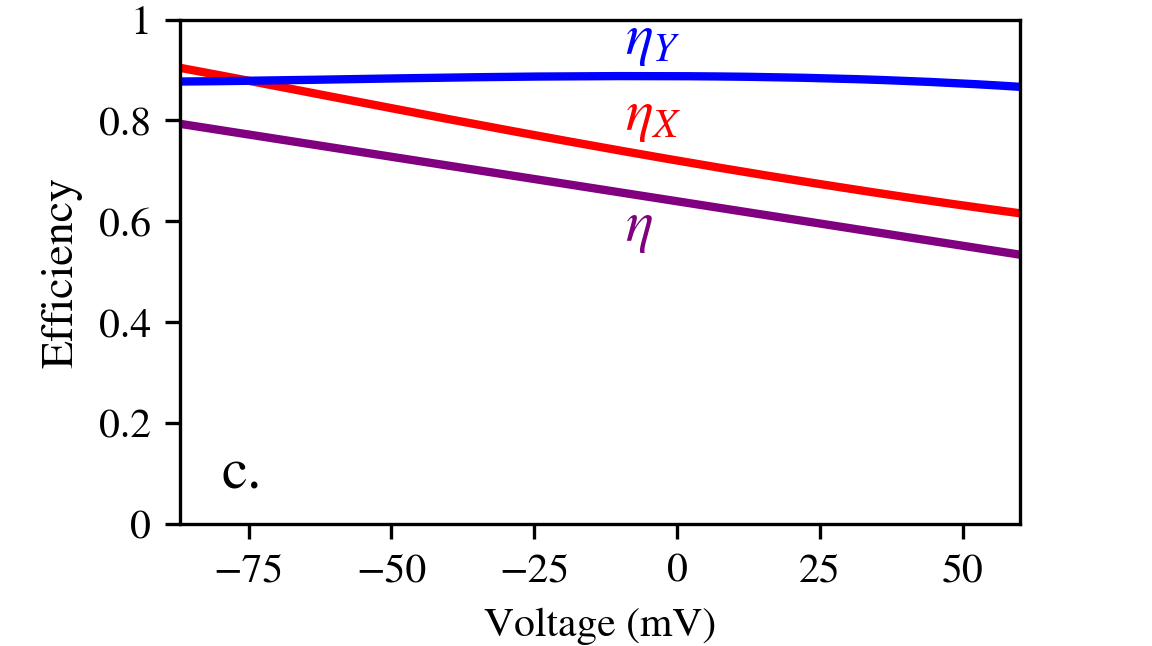}}
    \end{subfloat}
    \caption{Global probability current (a), subsystems' heat, internal energy, and information flow (b), and subsystems' efficiencies (c),  when varying voltage in a range encompassing the neuronal action potential. Environmental parameters taken from Table~\ref{tab:concentrations_neuron}.}
    \label{fig:voltage_variation}
\end{figure}

Figure~\ref{fig:voltage_turnover} shows that probability current increases with voltage. Figure~\ref{fig:voltage_energy_info} shows the internal information flow from the (ATP-consuming) $X$ subsystem to the (ion-transporting) $Y$ subsystem and the subsystems' heat flows when varying voltage. The information flow changes sign at $V=0.6\,\mathrm{mV}$, and subsystem $Y$'s heat flow changes sign at $V=-24.7\,\mathrm{mV}$. Therefore, for voltages higher than $-24.7\,\mathrm{mV}$ the subsystem $Y$ behaves as a conventional engine and not as a Maxwell demon, meaning that it does not exploit information in such a way that heat flows into either subsystem. Figure~\ref{fig:voltage_efficiencies} shows that the pump's global efficiency decreases with voltage, resulting primarily from the reduced efficiency of subsystem $X$. Overall, the pump continues to operate (positive net flux) despite the change in operational mode, however the efficiency decreases as the information flow decreases.

\section{\label{sec:discussion}Discussion}
In this paper, we studied the energy and information flows in cellular ion pumps. We used master-equation dynamics to find the nonequilibrium steady state for a pruned Albers-Post model for the sodium-potassium pump's main cycle. We then calculated the probability current, subsystem efficiencies, and internal energy and information flows. We considered the effect of variations in ion concentrations and also how changes in transmembrane voltage affect internal thermodynamics across a biologically relevant range including the neuronal action potential.

To analyze the energy and information flows we proposed a thermodynamically inspired bipartite partition. We found that the subsystem responsible for ATP hydrolysis and phosphorylation operates as a Maxwell demon, allowing for heat flow into the ion-transporting subsystem. Thus, a considerable information flow between the subsystems is required to satisfy the second law. Variation of external ion concentrations does not considerably impact the probability current, while variations of internal ion concentrations have a stronger effect. Finally, when increasing voltage over a range including the neuronal action potential, the probability current increases while global and subsystem efficiencies decrease. Notably, for higher voltages the Maxwell-demon behavior is lost, with both subsystems dissipating heat to the environment.

Our finding that the sodium-potassium pump relies on information flow as an important contribution to internal free energy transduction during its regular operation supports emerging evidence that information flows are widespread in biological molecular machines~\cite{Takaki2022_Information,leightonFlowEnergyInformation2025, leighton2024information}. This work reveals information flow in an entirely new class of systems, ion pumps, that perform important tasks in eukaryotic cells and consume a significant fraction of energy available for the cell. It also deepens our understanding of their thermodynamic behavior by providing an insight into the efficiencies of their internal dynamics.

Biophysically interpreting the presence of significant information flow remains an ongoing challenge for the field of information thermodynamics. In this particular case, the presence of significant information flow from subsystem $X$ to subsystem $Y$ suggests the ATP-hydrolysis subsystem senses changes in the ion binding/unbinding state, and acts quickly to rectify favorable fluctuations by changing the protein conformation and thus locking the bound ions in place. This would be achieved by protein conformational transition rates that depend strongly on the ion binding state such that transitions are both rapid and unlikely to reverse once the correct ion configuration is reached, and unlikely to occur otherwise. Such a set of transition rates comes at the cost of dissipated heat from ATP hydrolysis, which represents the thermodynamic cost of this information flow. The challenge of interpreting information flow more generally represents an important direction for future work.

Other future directions could include analyzing the general Albers-Post model for the sodium-potassium pump and using this methodology in different ion-transporting proteins such as the Cardiac Sarcoplasmic/Endoplasmic Ca$^{2+}$ (SERCA) pump~\cite{tranThermodynamicModelCardiac2009, toyoshimaStructuralBasisIon2004} or the Gastric Hydrogen Potassium ATPase~\cite{shinGastricHKATPaseStructure2009}, as well as extending it to efflux pumps~\cite{gerry2025specificity} or light-driven proton pumps like bacteriorhodopsin~\cite{pineroOptimizationNonequilibriumFree2024}. Work in this area would help us understand general principles of efficient ion-pumping molecular machines, and could aid in developing efficient artificial nanoscale pumps.

We noted that the definition of the bipartite decomposition is not unique; in this molecular machine, functional intuition led to a particular choice of partition, but this selection may not always be as straightforward. Our work highlights the importance of finding a general and consistent approach to define bipartite partitions on arbitrary systems. We also assumed the system to be in a nonequilibrium steady state, which is reasonable so long as environmental changes occur on slower timescales than the system's relaxation time. Future work could explore the thermodynamics of transient behavior under rapid environmental changes.

\begin{acknowledgments}
The authors thank Ronald Clarke (Department of Chemistry, University of Sydney) for helpful discussions. JDJP is grateful to Mitacs who sponsored his work through the Globalink Research Internship Program. This work was further supported by a Natural Sciences and Engineering Research Council of Canada (NSERC) CGS Doctoral fellowship (M.P.L.), a Mossman Postdoctoral Fellowship from the Department of Physics at Yale University (M.P.L.), an NSERC Discovery Grant and Discovery Accelerator Supplement RGPIN-2020-04950 (D.A.S.), an NSERC Alliance International Collaboration Grant ALLRP-585940-23 (D.A.S.), and a Tier-II Canada Research Chair CRC-2020-00098 (D.A.S.).
\end{acknowledgments}

\appendix

\section{Environmental parameters}\label{app:concentrations}

Tables \ref{tab:concentrations} and \ref{tab:concentrations_neuron} show the 
respective
resting-cell ion concentrations for a mammalian kidney cell and for a mammalian neuron at rest.

\begin{table}[h]
{\renewcommand{\arraystretch}{2}
\begin{tabular}{|l|r|}
\hline
{[}Na$^+${]}$_\text{in}$  & 15 mM           \\ \hline
{[}Na$^+${]}$_\text{out}$ & 140 mM          \\ \hline
{[}K$^+${]}$_\text{in}$   & 120 mM          \\ \hline
{[}K$^+${]}$_\text{out}$  & 4 mM            \\ \hline
{[}ATP{]}                 & 5 mM            \\ \hline
{[}ADP{]}                 & 0.1 mM          \\ \hline
{[}P$_{\rm i}${]}         & 5 mM            \\ \hline
$K_{\rm h}$               & 4$\cdot$10$^5$ mM \\ \hline
\end{tabular}
}
\caption{Different concentrations considered for the model taken from \cite{clarke_quantitative_2013}. $K_{\rm h}$ is the equilibrium constant for ATP hydrolysis.}
\label{tab:concentrations}
\end{table}

\begin{table}[h]
{\renewcommand{\arraystretch}{2}
\begin{tabular}{|l|r|}
\hline
{[}Na$^+${]}$_\text{in}$  & 12 mM           \\ \hline
{[}Na$^+${]}$_\text{out}$ & 145 mM          \\ \hline
{[}K$^+${]}$_\text{in}$   & 140 mM          \\ \hline
{[}K$^+${]}$_\text{out}$  & 5 mM            \\ \hline
{[}ATP{]}                 & 5 mM            \\ \hline
{[}ADP{]}                 & 0.1 mM          \\ \hline
{[}P$_{\rm i}${]}         & 5 mM            \\ \hline
$K_{\rm h}$               & 4$\cdot$10$^5$ mM \\ \hline
\end{tabular}
}
\caption{Different concentrations considered for resting neuron taken from \cite{tagluk_influence_2014}.}
\label{tab:concentrations_neuron}
\end{table}

\section{Model corrections}\label{app:addendum}
We made three changes to the model to correct minor typos in the original publication of Ref.~\cite{clarke_quantitative_2013}, which we confirmed with the corresponding author~\cite{private_communication}.

In Fig.~\ref{fig:Albers_Post}, the transition rate from state E$_1$Na$^+_2$ to state E$_1$Na$^+$ was changed from $k^i_{dN}$ to $2k^i_{dN}$. Also, we changed the value of coefficient $k^o_{bN1,V=0}$ from $10^3 M^{-1}s^{-1}$ to $10^6 M^{-1}s^{-1}$, and the value of coefficient $k^i_{bN1,V=0}$ from $1.67\cdot10^3 M^{-1}s^{-1}$ to $1.67\cdot 10^6 M^{-1}s^{-1}$.

\bibliography{references}

\end{document}

%% file: Albers-Post-Horizontal.tex
\tikzset{every picture/.style={line width=0.75pt}} 

\begin{tikzpicture}[x=0.75pt,y=0.75pt,yscale=-1,xscale=1]

\draw [line width=0.75]    (31.58,200.91) -- (31.72,108.78) ;
\draw [shift={(31.73,105.78)}, rotate = 90.09] [fill={rgb, 255:red, 0; green, 0; blue, 0 }  ][line width=0.08]  [draw opacity=0] (5.36,-2.57) -- (0,0) -- (5.36,2.57) -- cycle    ;
\draw [line width=0.75]    (39.64,199.2) -- (39.78,107.07) ;
\draw [shift={(39.63,202.2)}, rotate = 270.09] [fill={rgb, 255:red, 0; green, 0; blue, 0 }  ][line width=0.08]  [draw opacity=0] (5.36,-2.57) -- (0,0) -- (5.36,2.57) -- cycle    ;
\draw [line width=0.75]    (622.99,323.14) -- (623.13,240.89) ;
\draw [shift={(623.14,237.89)}, rotate = 90.1] [fill={rgb, 255:red, 0; green, 0; blue, 0 }  ][line width=0.08]  [draw opacity=0] (5.36,-2.57) -- (0,0) -- (5.36,2.57) -- cycle    ;
\draw [line width=0.75]    (631.05,321.3) -- (631.19,239.05) ;
\draw [shift={(631.04,324.3)}, rotate = 270.1] [fill={rgb, 255:red, 0; green, 0; blue, 0 }  ][line width=0.08]  [draw opacity=0] (5.36,-2.57) -- (0,0) -- (5.36,2.57) -- cycle    ;
\draw [line width=0.75]    (75.6,87.57) -- (127.64,87.71) ;
\draw [shift={(130.64,87.72)}, rotate = 180.15] [fill={rgb, 255:red, 0; green, 0; blue, 0 }  ][line width=0.08]  [draw opacity=0] (5.36,-2.57) -- (0,0) -- (5.36,2.57) -- cycle    ;
\draw [line width=0.75]    (78.61,79.31) -- (130.65,79.45) ;
\draw [shift={(75.61,79.3)}, rotate = 0.15] [fill={rgb, 255:red, 0; green, 0; blue, 0 }  ][line width=0.08]  [draw opacity=0] (5.36,-2.57) -- (0,0) -- (5.36,2.57) -- cycle    ;
\draw [color={rgb, 255:red, 128; green, 128; blue, 128 }  ,draw opacity=1 ][line width=0.75]    (294.44,330.99) -- (294.56,266.9) ;
\draw [shift={(294.57,263.9)}, rotate = 90.11] [fill={rgb, 255:red, 128; green, 128; blue, 128 }  ,fill opacity=1 ][line width=0.08]  [draw opacity=0] (5.36,-2.57) -- (0,0) -- (5.36,2.57) -- cycle    ;
\draw [color={rgb, 255:red, 128; green, 128; blue, 128 }  ,draw opacity=1 ][line width=0.75]    (302.5,328.9) -- (302.62,264.81) ;
\draw [shift={(302.49,331.9)}, rotate = 270.11] [fill={rgb, 255:red, 128; green, 128; blue, 128 }  ,fill opacity=1 ][line width=0.08]  [draw opacity=0] (5.36,-2.57) -- (0,0) -- (5.36,2.57) -- cycle    ;
\draw [color={rgb, 255:red, 128; green, 128; blue, 128 }  ,draw opacity=1 ][line width=0.75]    (510.46,331.46) -- (510.58,265) ;
\draw [shift={(510.58,262)}, rotate = 90.1] [fill={rgb, 255:red, 128; green, 128; blue, 128 }  ,fill opacity=1 ][line width=0.08]  [draw opacity=0] (5.36,-2.57) -- (0,0) -- (5.36,2.57) -- cycle    ;
\draw [color={rgb, 255:red, 128; green, 128; blue, 128 }  ,draw opacity=1 ][line width=0.75]    (518.51,329.41) -- (518.63,262.95) ;
\draw [shift={(518.51,332.41)}, rotate = 270.1] [fill={rgb, 255:red, 128; green, 128; blue, 128 }  ,fill opacity=1 ][line width=0.08]  [draw opacity=0] (5.36,-2.57) -- (0,0) -- (5.36,2.57) -- cycle    ;
\draw [color={rgb, 255:red, 128; green, 128; blue, 128 }  ,draw opacity=1 ][line width=0.75]    (301.87,167.04) -- (301.98,104) ;
\draw [shift={(301.99,101)}, rotate = 90.11] [fill={rgb, 255:red, 128; green, 128; blue, 128 }  ,fill opacity=1 ][line width=0.08]  [draw opacity=0] (5.36,-2.57) -- (0,0) -- (5.36,2.57) -- cycle    ;
\draw [color={rgb, 255:red, 128; green, 128; blue, 128 }  ,draw opacity=1 ][line width=0.75]    (292.26,164.9) -- (292.38,101.87) ;
\draw [shift={(292.25,167.9)}, rotate = 270.11] [fill={rgb, 255:red, 128; green, 128; blue, 128 }  ,fill opacity=1 ][line width=0.08]  [draw opacity=0] (5.36,-2.57) -- (0,0) -- (5.36,2.57) -- cycle    ;
\draw [color={rgb, 255:red, 128; green, 128; blue, 128 }  ,draw opacity=1 ][line width=0.75]    (169.74,194.89) -- (169.88,102.76) ;
\draw [shift={(169.73,197.89)}, rotate = 270.09] [fill={rgb, 255:red, 128; green, 128; blue, 128 }  ,fill opacity=1 ][line width=0.08]  [draw opacity=0] (5.36,-2.57) -- (0,0) -- (5.36,2.57) -- cycle    ;
\draw [color={rgb, 255:red, 128; green, 128; blue, 128 }  ,draw opacity=1 ][line width=0.75]    (172.75,324.4) -- (172.75,230.6) ;
\draw [shift={(172.75,327.4)}, rotate = 270] [fill={rgb, 255:red, 128; green, 128; blue, 128 }  ,fill opacity=1 ][line width=0.08]  [draw opacity=0] (5.36,-2.57) -- (0,0) -- (5.36,2.57) -- cycle    ;
\draw    (119.68,359.31) .. controls (120.67,351.79) and (112.51,346.55) .. (101.73,346.75) ;
\draw    (250.88,361.11) .. controls (251.87,353.59) and (243.71,348.35) .. (232.93,348.55) ;
\draw    (374.57,361.91) .. controls (375.56,354.39) and (367.4,349.15) .. (356.62,349.35) ;
\draw    (555.77,327.32) .. controls (554.78,334.83) and (562.94,340.07) .. (573.72,339.88) ;
\draw    (437.37,328.92) .. controls (436.38,336.43) and (444.54,341.67) .. (455.32,341.48) ;
\draw    (125.03,63.35) .. controls (125.99,72.99) and (117.81,79.68) .. (107.03,79.38) ;
\draw    (247.79,62.31) .. controls (248.76,71.95) and (240.57,78.63) .. (229.79,78.33) ;
\draw    (382.93,62.87) .. controls (383.89,72.51) and (375.71,79.2) .. (364.92,78.9) ;
\draw    (427.72,103.05) .. controls (426.76,93.85) and (434.94,87.46) .. (445.72,87.75) ;
\draw    (547.28,104.67) .. controls (546.32,95.46) and (554.5,89.07) .. (565.28,89.36) ;
\draw [color={rgb, 255:red, 128; green, 128; blue, 128 }  ,draw opacity=1 ]   (276.07,118.79) .. controls (285.71,117.83) and (292.4,126.01) .. (292.1,136.8) ;
\draw [color={rgb, 255:red, 128; green, 128; blue, 128 }  ,draw opacity=1 ]   (494.42,120.12) .. controls (504.06,119.16) and (510.75,127.34) .. (510.45,138.12) ;
\draw [color={rgb, 255:red, 128; green, 128; blue, 128 }  ,draw opacity=1 ]   (494.43,309.96) .. controls (504.06,310.92) and (510.78,303.04) .. (510.52,292.61) ;
\draw [color={rgb, 255:red, 128; green, 128; blue, 128 }  ,draw opacity=1 ]   (278.41,312.38) .. controls (288.04,313.35) and (294.76,305.46) .. (294.5,295.04) ;
\draw [line width=0.75]    (204.1,86.87) -- (256.14,87.01) ;
\draw [shift={(259.14,87.02)}, rotate = 180.15] [fill={rgb, 255:red, 0; green, 0; blue, 0 }  ][line width=0.08]  [draw opacity=0] (5.36,-2.57) -- (0,0) -- (5.36,2.57) -- cycle    ;
\draw [line width=0.75]    (207.11,78.61) -- (259.15,78.75) ;
\draw [shift={(204.11,78.6)}, rotate = 0.15] [fill={rgb, 255:red, 0; green, 0; blue, 0 }  ][line width=0.08]  [draw opacity=0] (5.36,-2.57) -- (0,0) -- (5.36,2.57) -- cycle    ;
\draw [line width=0.75]    (325.9,87.47) -- (377.94,87.61) ;
\draw [shift={(380.94,87.62)}, rotate = 180.15] [fill={rgb, 255:red, 0; green, 0; blue, 0 }  ][line width=0.08]  [draw opacity=0] (5.36,-2.57) -- (0,0) -- (5.36,2.57) -- cycle    ;
\draw [line width=0.75]    (328.91,79.21) -- (380.95,79.35) ;
\draw [shift={(325.91,79.2)}, rotate = 0.15] [fill={rgb, 255:red, 0; green, 0; blue, 0 }  ][line width=0.08]  [draw opacity=0] (5.36,-2.57) -- (0,0) -- (5.36,2.57) -- cycle    ;
\draw [line width=0.75]    (423,87.67) -- (475.04,87.81) ;
\draw [shift={(478.04,87.82)}, rotate = 180.15] [fill={rgb, 255:red, 0; green, 0; blue, 0 }  ][line width=0.08]  [draw opacity=0] (5.36,-2.57) -- (0,0) -- (5.36,2.57) -- cycle    ;
\draw [line width=0.75]    (426.01,79.41) -- (478.05,79.55) ;
\draw [shift={(423.01,79.4)}, rotate = 0.15] [fill={rgb, 255:red, 0; green, 0; blue, 0 }  ][line width=0.08]  [draw opacity=0] (5.36,-2.57) -- (0,0) -- (5.36,2.57) -- cycle    ;
\draw [line width=0.75]    (541.4,89.27) -- (593.44,89.41) ;
\draw [shift={(596.44,89.42)}, rotate = 180.15] [fill={rgb, 255:red, 0; green, 0; blue, 0 }  ][line width=0.08]  [draw opacity=0] (5.36,-2.57) -- (0,0) -- (5.36,2.57) -- cycle    ;
\draw [line width=0.75]    (544.41,81.01) -- (596.45,81.15) ;
\draw [shift={(541.41,81)}, rotate = 0.15] [fill={rgb, 255:red, 0; green, 0; blue, 0 }  ][line width=0.08]  [draw opacity=0] (5.36,-2.57) -- (0,0) -- (5.36,2.57) -- cycle    ;
\draw [line width=0.75]    (74.7,341.05) -- (126.74,340.91) ;
\draw [shift={(129.74,340.9)}, rotate = 179.85] [fill={rgb, 255:red, 0; green, 0; blue, 0 }  ][line width=0.08]  [draw opacity=0] (5.36,-2.57) -- (0,0) -- (5.36,2.57) -- cycle    ;
\draw [line width=0.75]    (77.21,346.81) -- (129.25,346.67) ;
\draw [shift={(74.21,346.82)}, rotate = 359.85] [fill={rgb, 255:red, 0; green, 0; blue, 0 }  ][line width=0.08]  [draw opacity=0] (5.36,-2.57) -- (0,0) -- (5.36,2.57) -- cycle    ;
\draw [line width=0.75]    (205.4,340.35) -- (257.44,340.21) ;
\draw [shift={(260.44,340.2)}, rotate = 179.85] [fill={rgb, 255:red, 0; green, 0; blue, 0 }  ][line width=0.08]  [draw opacity=0] (5.36,-2.57) -- (0,0) -- (5.36,2.57) -- cycle    ;
\draw [line width=0.75]    (208.41,348.61) -- (260.45,348.47) ;
\draw [shift={(205.41,348.62)}, rotate = 359.85] [fill={rgb, 255:red, 0; green, 0; blue, 0 }  ][line width=0.08]  [draw opacity=0] (5.36,-2.57) -- (0,0) -- (5.36,2.57) -- cycle    ;
\draw [line width=0.75]    (329.1,341.15) -- (381.14,341.01) ;
\draw [shift={(384.14,341)}, rotate = 179.85] [fill={rgb, 255:red, 0; green, 0; blue, 0 }  ][line width=0.08]  [draw opacity=0] (5.36,-2.57) -- (0,0) -- (5.36,2.57) -- cycle    ;
\draw [line width=0.75]    (332.11,349.41) -- (384.15,349.27) ;
\draw [shift={(329.11,349.42)}, rotate = 359.85] [fill={rgb, 255:red, 0; green, 0; blue, 0 }  ][line width=0.08]  [draw opacity=0] (5.36,-2.57) -- (0,0) -- (5.36,2.57) -- cycle    ;
\draw [line width=0.75]    (427.8,341.55) -- (479.84,341.41) ;
\draw [shift={(482.84,341.4)}, rotate = 179.85] [fill={rgb, 255:red, 0; green, 0; blue, 0 }  ][line width=0.08]  [draw opacity=0] (5.36,-2.57) -- (0,0) -- (5.36,2.57) -- cycle    ;
\draw [line width=0.75]    (430.81,349.81) -- (482.85,349.67) ;
\draw [shift={(427.81,349.82)}, rotate = 359.85] [fill={rgb, 255:red, 0; green, 0; blue, 0 }  ][line width=0.08]  [draw opacity=0] (5.36,-2.57) -- (0,0) -- (5.36,2.57) -- cycle    ;
\draw [line width=0.75]    (546.2,339.95) -- (598.24,339.81) ;
\draw [shift={(601.24,339.8)}, rotate = 179.85] [fill={rgb, 255:red, 0; green, 0; blue, 0 }  ][line width=0.08]  [draw opacity=0] (5.36,-2.57) -- (0,0) -- (5.36,2.57) -- cycle    ;
\draw [line width=0.75]    (549.21,348.21) -- (601.25,348.07) ;
\draw [shift={(546.21,348.22)}, rotate = 359.85] [fill={rgb, 255:red, 0; green, 0; blue, 0 }  ][line width=0.08]  [draw opacity=0] (5.36,-2.57) -- (0,0) -- (5.36,2.57) -- cycle    ;
\draw [color={rgb, 255:red, 128; green, 128; blue, 128 }  ,draw opacity=1 ][line width=0.75]    (519.87,170.58) -- (519.99,103.6) ;
\draw [shift={(519.99,100.6)}, rotate = 90.1] [fill={rgb, 255:red, 128; green, 128; blue, 128 }  ,fill opacity=1 ][line width=0.08]  [draw opacity=0] (5.36,-2.57) -- (0,0) -- (5.36,2.57) -- cycle    ;
\draw [color={rgb, 255:red, 128; green, 128; blue, 128 }  ,draw opacity=1 ][line width=0.75]    (510.26,168.5) -- (510.38,101.52) ;
\draw [shift={(510.25,171.5)}, rotate = 270.1] [fill={rgb, 255:red, 128; green, 128; blue, 128 }  ,fill opacity=1 ][line width=0.08]  [draw opacity=0] (5.36,-2.57) -- (0,0) -- (5.36,2.57) -- cycle    ;
\draw [line width=0.75]    (32.88,324.17) -- (33.16,239.09) ;
\draw [shift={(33.17,236.09)}, rotate = 90.19] [fill={rgb, 255:red, 0; green, 0; blue, 0 }  ][line width=0.08]  [draw opacity=0] (5.36,-2.57) -- (0,0) -- (5.36,2.57) -- cycle    ;
\draw    (23.01,267.59) .. controls (34.59,261.25) and (35.77,248.8) .. (25.3,242.05) ;
\draw [shift={(22.74,240.63)}, rotate = 25.46] [fill={rgb, 255:red, 0; green, 0; blue, 0 }  ][line width=0.08]  [draw opacity=0] (3.57,-1.72) -- (0,0) -- (3.57,1.72) -- cycle    ;
\draw [color={rgb, 255:red, 255; green, 0; blue, 0 }  ,draw opacity=1 ][line width=0.75]    (41.97,321.4) -- (42.25,236.31) ;
\draw [shift={(41.96,324.4)}, rotate = 270.19] [fill={rgb, 255:red, 255; green, 0; blue, 0 }  ,fill opacity=1 ][line width=0.08]  [draw opacity=0] (5.36,-2.57) -- (0,0) -- (5.36,2.57) -- cycle    ;
\draw [line width=0.75]    (631.86,196.92) -- (631.62,111.41) ;
\draw [shift={(631.86,199.92)}, rotate = 269.84] [fill={rgb, 255:red, 0; green, 0; blue, 0 }  ][line width=0.08]  [draw opacity=0] (5.36,-2.57) -- (0,0) -- (5.36,2.57) -- cycle    ;
\draw    (631.76,161.98) .. controls (630.94,167.57) and (632.98,171.96) .. (638.08,174.02) ;
\draw [shift={(640.94,174.86)}, rotate = 191.4] [fill={rgb, 255:red, 0; green, 0; blue, 0 }  ][line width=0.08]  [draw opacity=0] (3.57,-1.72) -- (0,0) -- (3.57,1.72) -- cycle    ;
\draw [color={rgb, 255:red, 255; green, 0; blue, 0 }  ,draw opacity=1 ][line width=0.75]    (622.78,198.17) -- (622.54,115.11) ;
\draw [shift={(622.54,112.11)}, rotate = 89.84] [fill={rgb, 255:red, 255; green, 0; blue, 0 }  ,fill opacity=1 ][line width=0.08]  [draw opacity=0] (5.36,-2.57) -- (0,0) -- (5.36,2.57) -- cycle    ;

\draw (13.68,331.98) node [anchor=north west][inner sep=0.75pt]  [font=\normalsize]  {$\text{E}_{1}\text{Na}_{3}^{+}$};
\draw (10.63,206.37) node [anchor=north west][inner sep=0.75pt]  [font=\normalsize]  {$\text{E}_{1}\text{P}\left(\text{Na}^{+}\right)_{3}$};
\draw (11.38,73.29) node [anchor=north west][inner sep=0.75pt]  [font=\normalsize]  {$\text{E}_{2}\text{PNa}_{3}^{+}$};
\draw (140.93,72.99) node [anchor=north west][inner sep=0.75pt]  [font=\normalsize]  {$\text{E}_{2}\text{PNa}_{2}^{+}$};
\draw (268.38,70.58) node [anchor=north west][inner sep=0.75pt]  [font=\normalsize]  {$\text{E}_{2}\text{PNa}^{+}$};
\draw (388.83,75.03) node [anchor=north west][inner sep=0.75pt]  [font=\normalsize]  {$\text{E}_{2}\text{P}$};
\draw (491.9,72.53) node [anchor=north west][inner sep=0.75pt]  [font=\normalsize]  {$\text{E}_{2}\text{PK}^{+}$};
\draw (603.45,71.31) node [anchor=north west][inner sep=0.75pt]  [font=\normalsize]  {$\text{E}_{2}\text{PK}_{2}^{+}$};
\draw (595.28,208.7) node [anchor=north west][inner sep=0.75pt]  [font=\normalsize]  {$\text{E}_{2}\left(\text{K}^{+}\right)_{2}$};
\draw (610.29,331.27) node [anchor=north west][inner sep=0.75pt]  [font=\normalsize]  {$\text{E}_{1}\text{K}_{2}^{+}$};
\draw (498.86,333.12) node [anchor=north west][inner sep=0.75pt]  [font=\normalsize]  {$\text{E}_{1}\text{K}^{+}$};
\draw (398.37,335.97) node [anchor=north west][inner sep=0.75pt]  [font=\normalsize]  {$\text{E}_{1}$};
\draw (276.45,331.25) node [anchor=north west][inner sep=0.75pt]  [font=\normalsize]  {$\text{E}_{1}\text{Na}^{+}$};
\draw (148.97,329.68) node [anchor=north west][inner sep=0.75pt]  [font=\normalsize]  {$\text{E}_{1}\text{Na}_{2}^{+}$};
\draw (138.71,204.36) node [anchor=north west][inner sep=0.75pt]  [font=\normalsize,color={rgb, 255:red, 128; green, 128; blue, 128 }  ,opacity=1 ]  {$\text{E}_{2}\left(\text{Na}^{+}\right)_{2}$};
\draw (483.62,235.01) node [anchor=north west][inner sep=0.75pt]  [font=\normalsize,color={rgb, 255:red, 128; green, 128; blue, 128 }  ,opacity=1 ]  {$\text{E}_{1}\text{K}^{+}\text{Na}^{+}$};
\draw (264.46,236.22) node [anchor=north west][inner sep=0.75pt]  [font=\normalsize,color={rgb, 255:red, 128; green, 128; blue, 128 }  ,opacity=1 ]  {$\text{E}_{1}\text{Na}^{+}\text{K}^{+}$};
\draw (261.19,172.1) node [anchor=north west][inner sep=0.75pt]  [font=\normalsize,color={rgb, 255:red, 128; green, 128; blue, 128 }  ,opacity=1 ]  {$\text{E}_{2}\text{PNa}^{+}\text{K}^{+}$};
\draw (478.52,172.44) node [anchor=north west][inner sep=0.75pt]  [font=\normalsize,color={rgb, 255:red, 128; green, 128; blue, 128 }  ,opacity=1 ]  {$\text{E}_{2}\text{PK}^{+}\text{Na}^{+}$};
\draw (11.49,165.19) node [anchor=north west][inner sep=0.75pt]  [rotate=-270.11]  {$k_{2f}$};
\draw (41.68,165.92) node [anchor=north west][inner sep=0.75pt]  [rotate=-270.11]  {$k_{2r}$};
\draw (81.44,54.55) node [anchor=north west][inner sep=0.75pt]    {$k_{b\text{N1}}^{o}$};
\draw (81.87,92.59) node [anchor=north west][inner sep=0.75pt]    {$k_{d\text{N1}}^{o}$};
\draw (203.27,55.56) node [anchor=north west][inner sep=0.75pt]    {$k_{b\text{N}}^{o}$};
\draw (199.31,92.07) node [anchor=north west][inner sep=0.75pt]    {$2k_{d\text{N}}^{o}$};
\draw (440.87,90.48) node [anchor=north west][inner sep=0.75pt]    {$2k_{b\text{K}}^{o}$};
\draw (443.62,54.63) node [anchor=north west][inner sep=0.75pt]    {$k_{d\text{K}}^{o}$};
\draw (74.21,352.72) node [anchor=north west][inner sep=0.75pt]    {$k_{b\text{N1}}^{i}$};
\draw (78.64,314.86) node [anchor=north west][inner sep=0.75pt]    {$k_{d\text{N1}}^{i}$};
\draw (327.61,354.61) node [anchor=north west][inner sep=0.75pt]    {$2k_{b\text{N}}^{i}$};
\draw (335.85,317.92) node [anchor=north west][inner sep=0.75pt]    {$k_{d\text{N}}^{i}$};
\draw (579.05,316.52) node [anchor=north west][inner sep=0.75pt]    {$k_{b\text{K}}^{i}$};
\draw (571.82,355.64) node [anchor=north west][inner sep=0.75pt]    {$2k_{d\text{K}}^{i}$};
\draw (340.61,90.99) node [anchor=north west][inner sep=0.75pt]    {$k_{d\text{N}}^{o}$};
\draw (332.27,54.81) node [anchor=north west][inner sep=0.75pt]    {$2k_{b\text{N}}^{o}$};
\draw (573.68,96.21) node [anchor=north west][inner sep=0.75pt]    {$k_{b\text{K}}^{o}$};
\draw (568.42,56.79) node [anchor=north west][inner sep=0.75pt]    {$2k_{d\text{K}}^{o}$};
\draw (631.29,290.71) node [anchor=north west][inner sep=0.75pt]  [rotate=-270.11]  {$k_{4f}$};
\draw (603.52,290.96) node [anchor=north west][inner sep=0.75pt]  [rotate=-270.11]  {$k_{4r}$};
\draw (454.86,316.96) node [anchor=north west][inner sep=0.75pt]    {$2k_{b\text{K}}^{i}$};
\draw (458.13,354.32) node [anchor=north west][inner sep=0.75pt]    {$k_{d\text{K}}^{i}$};
\draw (215.18,315.19) node [anchor=north west][inner sep=0.75pt]    {$2k_{d\text{N}}^{i}$};
\draw (205.41,356.02) node [anchor=north west][inner sep=0.75pt]    {$k_{b\text{N}}^{i}$};
\draw (146.32,155.12) node [anchor=north west][inner sep=0.75pt]  [color={rgb, 255:red, 128; green, 128; blue, 128 }  ,opacity=1 ,rotate=-270.11]  {$k_{31}$};
\draw (148.09,290.01) node [anchor=north west][inner sep=0.75pt]  [color={rgb, 255:red, 128; green, 128; blue, 128 }  ,opacity=1 ,rotate=-270.11]  {$k_{4f}$};
\draw (268.77,295.09) node [anchor=north west][inner sep=0.75pt]  [color={rgb, 255:red, 128; green, 128; blue, 128 }  ,opacity=1 ,rotate=-270.11]  {$k_{b\text{K}}^{i}$};
\draw (305.96,296.36) node [anchor=north west][inner sep=0.75pt]  [color={rgb, 255:red, 128; green, 128; blue, 128 }  ,opacity=1 ,rotate=-270.11]  {$k_{d\text{K}}^{i}$};
\draw (518.06,291.4) node [anchor=north west][inner sep=0.75pt]  [color={rgb, 255:red, 128; green, 128; blue, 128 }  ,opacity=1 ,rotate=-270.11]  {$k_{d\text{N}}^{i}$};
\draw (481.07,291.34) node [anchor=north west][inner sep=0.75pt]  [color={rgb, 255:red, 128; green, 128; blue, 128 }  ,opacity=1 ,rotate=-270.11]  {$k_{b\text{N}}^{i}$};
\draw (304.01,159.13) node [anchor=north west][inner sep=0.75pt]  [color={rgb, 255:red, 128; green, 128; blue, 128 }  ,opacity=1 ,rotate=-270.11]  {$k_{d\text{K}}^{o}$};
\draw (266.35,157.46) node [anchor=north west][inner sep=0.75pt]  [color={rgb, 255:red, 128; green, 128; blue, 128 }  ,opacity=1 ,rotate=-270.11]  {$k_{b\text{K}}^{o}$};
\draw (483.6,161.62) node [anchor=north west][inner sep=0.75pt]  [color={rgb, 255:red, 128; green, 128; blue, 128 }  ,opacity=1 ,rotate=-270.11]  {$k_{b\text{N}}^{o}$};
\draw (519.14,159.14) node [anchor=north west][inner sep=0.75pt]  [color={rgb, 255:red, 128; green, 128; blue, 128 }  ,opacity=1 ,rotate=-270.11]  {$k_{d\text{N}}^{o}$};
\draw (115.16,359.37) node [anchor=north west][inner sep=0.75pt]    {$\text{Na}^{+}$};
\draw (243.91,362.35) node [anchor=north west][inner sep=0.75pt]    {$\text{Na}^{+}$};
\draw (362.58,360.07) node [anchor=north west][inner sep=0.75pt]    {$\text{Na}^{+}$};
\draw (426.79,306.14) node [anchor=north west][inner sep=0.75pt]    {$\text{K}^{+}$};
\draw (547.25,303.2) node [anchor=north west][inner sep=0.75pt]    {$\text{K}^{+}$};
\draw (113.36,43.7) node [anchor=north west][inner sep=0.75pt]    {$\text{Na}^{+}$};
\draw (233.91,42.33) node [anchor=north west][inner sep=0.75pt]    {$\text{Na}^{+}$};
\draw (371.08,43.15) node [anchor=north west][inner sep=0.75pt]    {$\text{Na}^{+}$};
\draw (414.59,101.65) node [anchor=north west][inner sep=0.75pt]    {$\text{K}^{+}$};
\draw (537.53,102.55) node [anchor=north west][inner sep=0.75pt]    {$\text{K}^{+}$};
\draw (255.01,106.83) node [anchor=north west][inner sep=0.75pt]  [color={rgb, 255:red, 128; green, 128; blue, 128 }  ,opacity=1 ]  {$\text{K}^{+}$};
\draw (468.19,107.83) node [anchor=north west][inner sep=0.75pt]  [color={rgb, 255:red, 128; green, 128; blue, 128 }  ,opacity=1 ]  {$\text{Na}^{+}$};
\draw (469.38,295.15) node [anchor=north west][inner sep=0.75pt]  [color={rgb, 255:red, 128; green, 128; blue, 128 }  ,opacity=1 ]  {$\text{Na}^{+}$};
\draw (258.01,300.55) node [anchor=north west][inner sep=0.75pt]  [color={rgb, 255:red, 128; green, 128; blue, 128 }  ,opacity=1 ]  {$\text{K}^{+}$};
\draw (10.46,319.58) node [anchor=north west][inner sep=0.75pt]  [rotate=-270.22]  {$k_{1f}$};
\draw (9.38,289.44) node [anchor=north west][inner sep=0.75pt]  [rotate=-270.22] [align=left] {ATP};
\draw (9.01,259.57) node [anchor=north west][inner sep=0.75pt]  [rotate=-270.22] [align=left] {ADP};
\draw (41.7,320.1) node [anchor=north west][inner sep=0.75pt]  [color={rgb, 255:red, 255; green, 0; blue, 0 }  ,opacity=1 ,rotate=-270.22]  {$k_{1r}$};
\draw (635.13,153.65) node [anchor=north west][inner sep=0.75pt]  [rotate=-269.84]  {$k_{32f}$};
\draw (642.5,163.81) node [anchor=north west][inner sep=0.75pt]    {$\text{P}_{i}$};
\draw (601.39,151.95) node [anchor=north west][inner sep=0.75pt]  [color={rgb, 255:red, 255; green, 0; blue, 0 }  ,opacity=1 ,rotate=-269.84]  {$k_{32r}$};

\end{tikzpicture}